\documentclass[10pt,conference]{IEEEtran}
\IEEEoverridecommandlockouts

\usepackage{booktabs}
\usepackage{xcolor}
\usepackage{soul} 
\usepackage{makecell}
\usepackage{xspace}
\usepackage{listings}
\usepackage{hyperref}
\usepackage{subcaption}
\usepackage{soul}
\setulcolor{blue}

\usepackage{balance}

\usepackage[utf8]{inputenc}
\usepackage{csvsimple}
\usepackage{booktabs}
\usepackage{array}
\usepackage{adjustbox}

\newcommand{\js}{{Java\-Script}\xspace}

\newcommand{\code}[1]{\text{\lstinline[basicstyle=\ttfamily\small, language=Java]~#1~}}

\definecolor{light_green}{rgb}{0.3, 0.7, 0.33}
\definecolor{light_red}{rgb}{0.7, 0.32, 0.31}

\definecolor{sh_comment}{rgb}{0.12, 0.38, 0.18}
\definecolor{sh_keyword}{rgb}{0.37, 0.08, 0.25}  
\definecolor{sh_string}{rgb}{0.06, 0.10, 0.98} 
\def\lstsmallmath{\leavevmode\ifmmode \scriptstyle \else  \fi}
\def\lstsmallmathend{\leavevmode\ifmmode  \else  \fi}
\definecolor{KWColor}{rgb}{0.5,0,0.67}
\definecolor{CommentColor}{rgb}{0.15,0.5,0.15}
\definecolor{lightgrey}{rgb}{0.8,0.8,0.8}

\lstdefinelanguage{JavaScript}[]{Java}{
   morekeywords={var,class,object,function, async, await, undefined, let, form, button, div, useState, number, textarea, then} 
} 

\lstdefinestyle{Eclipse}{
  xleftmargin=0pt,
  language = JavaScript,
  basicstyle=\sffamily\footnotesize,
  stringstyle=\color{sh_string},
  keywordstyle = \color{sh_keyword}\bfseries,  
  lineskip=-0.0em,
  commentstyle=\color{sh_comment}\itshape,  
  escapeinside={/*@}{@*/},
  numbersep=5pt,
  captionpos=b,
  xleftmargin=0.4cm, xrightmargin=0.5cm,
   morekeywords={invokestatic,invokeinterface,invokevirtual,invokespecial,then},
}

\hypersetup{
  colorlinks=true,
  linkcolor=blue,
  urlcolor=blue,
  citecolor=blue
}

\newcommand\numOfProjects{30\xspace}
\newcommand\numOfFiles{16{,}060\xspace}
\newcommand\numOfFunctions{336{,}710\xspace}
\newcommand\totalLOC{42.5M\xspace}

\newcommand\medianIQR{0.14\xspace}

\newcommand\minMedianCTS{0.77\xspace} 
\newcommand\maxMedianCTS{0.93\xspace} 

\usepackage{cite}
\usepackage{amsmath,amssymb,amsfonts}
\usepackage{algorithmic}
\usepackage{graphicx}
\usepackage{textcomp}
\usepackage{xcolor}
\def\BibTeX{{\rm B\kern-.05em{\sc i\kern-.025em b}\kern-.08em
    T\kern-.1667em\lower.7ex\hbox{E}\kern-.125emX}}
\begin{document}

\title{Characterizing Structural Testability in \js:\\An Empirical Study}

\author{\IEEEauthorblockN{Anonymous Author(s)}}

\author{\IEEEauthorblockN{Shahrzad Mirzaei\IEEEauthorrefmark{1}}
\IEEEauthorblockA{
\textit{Motorola Solutions}\\
BC, Canada \\
shahrzad\_mirzaei@sfu.ca}
\thanks{\IEEEauthorrefmark{1} Work reported in this paper was conducted while pursuing a Master's degree at Simon Fraser University.}

\and

\IEEEauthorblockN{Saba Alimadadi}
\IEEEauthorblockA{
\textit{Simon Fraser University}\\
BC, Canada \\
saba@sfu.ca}
}

\maketitle

\begin{abstract}

Software testability has long been recognized as a software quality attribute that influences testing effort and effectiveness.
While prior work has extensively studied testability in object-oriented systems and, more recently, concurrent software, comparatively little is known about how structural testability manifests in modern \js ecosystems. \js applications rely heavily on asynchronous execution, event-driven control flow, closures, and dynamic interactions, yet these constructs are not explicitly captured by existing testability frameworks.

This paper presents a large-scale empirical study of structural testability in \js systems. To enable this study, we operationalize structural testability as a seven-dimensional construct capturing controllability, observability, branching complexity, asynchronous coordination, event-driven behaviour, encapsulation, and side-effect intensity. These dimensions are derived from AST-based static analysis and aggregated into a Composite Testability Score (CTS) that supports comparative analysis across functions, files, and projects.

We apply this framework to 30 open-source \js projects spanning diverse domains and sizes. Our analysis characterizes the distribution of structural testability, identifies recurring structural archetypes among low-CTS functions, and examines associations between project characteristics and testability patterns. We find that structurally-challenging functions are concentrated within a relatively small subset of files and emerge through multiple recurring structural configurations rather than a single dominant pattern. At the ecosystem level, project activity and size exhibit stronger associations with structural testability than project age, popularity, or framework choice.
These findings provide new insight into how structural testability manifests in modern \js systems and establish a foundation for future studies of testing effort, test generation effectiveness, testability-aware refactoring, and software quality assessment in dynamic and event-driven environments.

\end{abstract}

\begin{IEEEkeywords}
software testability, \js, static analysis, empirical software engineering, software metrics.
\end{IEEEkeywords}

\section{Introduction}
\label{sec:introduction}

Software testability has long been recognized as an important software quality attribute. ISO/IEC defines testability in terms of the effectiveness and efficiency with which testing criteria can be established and testing activities can be performed~\cite{iso25000}.
Foundational work by Freedman~\cite{freedman:testability} and Binder~\cite{binder1994design,binder:oo} identifies controllability and observability as key factors influencing testing effectiveness. Complementary perspectives, such as the PIE model of Voas and Miller~\cite{voas:testability}, relate testability to a program's ability to reveal existing faults during testing. Collectively, these studies establish testability as a multifaceted software quality attribute that influences the effort and effectiveness of software testing.

Among these perspectives, a substantial body of research has focused on \emph{structural testability}, studying how measurable software characteristics influence testing effort and effectiveness. Prior work has examined relationships between testability and structural properties such as complexity, coupling, cohesion, dependencies,  encapsulation, and object-oriented characteristics~\cite{bruntink:testability,mouchawrab2005measurement,badri2011empirical,jungmayr:testability,terragni:testability,garousi:survey}.
These studies have largely focused on statically-typed, object-oriented systems, particularly Java and C++, where testability has been investigated at the method, class, component, and system levels.
More recently, researchers have extended classical testability concepts to concurrent systems through metrics capturing synchronization, shared-state interactions, and execution coordination~\cite{yu2016predicting}. However, comparatively little is known about the structural testability characteristics of modern \js ecosystems.

\js is among the most widely-deployed programming languages, powering modern web front-ends, server-side systems through Node.js, and large-scale developer tooling ecosystems. Its execution model differs substantially from the object-oriented systems that dominate the testability literature. Modern \js applications rely heavily on asynchronous execution, event-driven control flow, closures, dynamic module interactions, and framework-specific coordination patterns~\cite{andreasen:survey}.
Prior work has shown that several of these characteristics complicate testing and automated test generation. For example, Fard and Mesbah~\cite{fard:uncovered} report that closures, callbacks, and dynamic DOM interactions frequently impede test coverage, while subsequent studies identify asynchronous coordination as a recurring source of testing difficulty, flakiness, and inadequacy~\cite{alimadadi:promises,hashemi:flaky,arteca:nessie,turcotte:drasync,selakovic:higherorder,ganji:jscope}. Despite these observations, existing measurement approaches largely rely on traditional proxy metrics such as cyclomatic complexity, coupling, or LOC~\cite{carstensen:thesis}, which do not explicitly capture many structural characteristics central to modern \js systems.

This gap motivates the need for a language-aware operationalization of structural testability for \js. Rather than introducing a new theory of testability, our goal is to adapt established testability dimensions to the structural characteristics of contemporary \js ecosystems and empirically characterize their distribution at scale. In particular, we operationalize seven function-level dimensions capturing controllability, observability, branching complexity, encapsulation, side effects, asynchronous coordination, and event-driven execution. These dimensions are computed entirely through AST-based static analysis and aggregated into a composite testability score (CTS) that supports comparative analysis across functions, files, and projects.

Using this framework, we conduct a large-scale empirical study of 30 open-source \js projects comprising 337K functions and 42.5M lines of code. Our study addresses three research questions. RQ1 characterizes the distribution of structural testability across functions, files, and projects. RQ2 examines the structural characteristics and recurring profiles of low-CTS functions. RQ3 investigates how project characteristics and testing practices relate to structural testability patterns.

Our findings reveal several consistent structural patterns. First, low-CTS functions exhibit strong structural concentration: a relatively small fraction of files accounts for a disproportionate share of structurally-challenging code. Second, low-CTS functions are not homogeneous. Clustering analysis identifies multiple recurring structural archetypes, including asynchronous coordination routines, side-effect-dominated code, and structurally-unobservable functions.
Finally, structural testability exhibits measurable associations with project-level characteristics, suggesting that structurally-challenging code emerges within broader ecosystem and architectural contexts rather than solely through isolated code-level decisions.

These findings contribute to a growing body of research on software testability by extending structural testability analysis to modern \js ecosystems. Beyond improving our understanding of \js systems, the resulting framework provides a foundation for future studies of testing effort, automated test generation, testability-aware refactoring, and software quality assessment in dynamic and event-driven environments.

This paper makes the following contributions:

\begin{itemize}
\item A \js-aware operationalization of established structural testability dimensions, extending classical testability models with measures for asynchronous coordination and event-driven execution.

\item A fully automated static-analysis framework that computes and aggregates seven structural testability dimensions across function, file, and project levels.

\item A large-scale empirical characterization of structural testability across 30 real-world JavaScript projects comprising 337K functions and 42.5M LOC.

\item An open-source implementation of the analysis framework, together with the experimental pipeline and datasets, released to support reproducibility and future research.
\footnote{\url{https://github.com/SEatSFU/JSTestabilityStudy}}

\end{itemize}

\section{\js Testability Metrics}
\label{sec:metrics}

This section introduces the structural testability metrics used in our study. 
The metrics are designed to capture properties of \js code that influence the ease of controlling execution and observing behaviour during testing. 
All metrics are computed statically at the function level and later aggregated to file and project levels.

\subsection{Theoretical Foundations}
\label{subsec:theoretical-foundations}

Software testability has traditionally been characterized through dimensions such as controllability and observability~\cite{freedman:testability,binder1994design,binder:oo}.
ISO/IEC 25010 defines testability as the degree of effectiveness and efficiency with which test criteria can be established and tests can be performed to determine whether those criteria have been met~\cite{iso25000}.
Complementary perspectives such as Voas and Miller's PIE model relate testability to fault revelation during testing~\cite{voas:testability}.

Building on these foundations, prior work operationalized testability using structural software metrics, particularly for object-oriented systems, examining factors such as complexity, dependencies, encapsulation, cohesion, coupling, inheritance, and polymorphism~\cite{bruntink:testability,mouchawrab2005measurement,badri2011empirical,jungmayr:testability,mcgregor1996measure}.
More recently, researchers have extended classical testability concepts to concurrent and asynchronous systems through metrics capturing synchronization, shared-state interactions, and execution coordination~\cite{yu2016predicting}. 

Our work follows this tradition by operationalizing established testability dimensions for modern \js ecosystems. Specifically, we adapt classical notions of controllability, observability, complexity, encapsulation, and side effects to \js-specific constructs such as closures and module-based architectures, while extending prior models with dimensions for asynchronous coordination and event-driven execution.
Following prior structural testability and software quality studies, we also employ a composite indicator to support comparative characterization across software artifacts while preserving dimension-level analyses for interpretation.


\subsection{Design Goals}
\label{subsec:metric-goals}

The metrics are guided by four goals:

\begin{itemize}
  \item Language-aware operationalization: adapt established testability dimensions to \js-specific constructs such as asynchronous execution, event-driven programming, closures, and module boundaries.
  \item Static computability: rely solely on static analysis without requiring test execution or runtime traces.
  \item Actionability: reflect structural properties that developers can modify through refactoring or design changes.
  \item Empirical suitability: scale to large repositories and support aggregation across functions, files, and projects.
\end{itemize}

\subsection{Metric Definitions}
\label{subsec:metric-definitions}

Following prior testability literature, we operationalize seven function-level dimensions capturing controllability, observability, structural complexity, encapsulation, asynchronous behaviour, event-driven execution, and side effects.

Five of the seven dimensions (C, O, BC, EL, and SEI) adapt established structural testability and software-quality constructs to \js. The remaining two dimensions (AC and EC) extend classical testability models to capture asynchronous coordination and event-driven execution patterns that are central to modern \js systems but largely absent from earlier testability frameworks.

\subsubsection{Controllability (C)}
\label{subsubsec:controllability}

Controllability measures how easily a test can influence a function’s execution through explicit inputs. 
It increases with exposed parameters and exported interfaces, and decreases with hidden dependencies such as global variables, closure-captured state, or direct instantiation of imported modules.

\vspace{-12pt}

\begin{small}

\begin{multline*}
C = w_P \cdot \#params + w_D \cdot \#defaultParams + w_E \cdot isExported \\
    - w_G \cdot \#globalRefs - w_{Cl} \cdot \#closureRefs \\
    - w_{SI} \cdot \#staticImportsNewables
\end{multline*}

\end{small}

This metric operationalizes the classical notion of controllability~\cite{freedman:testability,binder:oo,mcgregor1996measure} in the context of \js by explicitly accounting for closure-captured state, global dependencies, and module interactions.

\subsubsection{Observability (O)}
\label{subsubsec:observability}

Observability is one of the foundational dimensions of software testability~\cite{freedman:testability,mcgregor1996measure}, reflecting how easily a test can observe the effects of executing a function. 
Observable outputs include return values, logging, or explicit I/O operations, while mutations to hidden state reduce observability.

\vspace{-14pt}

\begin{small}

\begin{multline*}
O = w_R \cdot hasReturnValue + w_L \cdot \#logWrites \\
    + w_{IO} \cdot \#ioWrites - w_H \cdot \#hiddenStateWrites
\end{multline*}

\end{small}


This metric operationalizes the classical notion of observability~\cite{freedman:testability} for \js programs by explicitly recognizing common observable effects beyond return values, while accounting for hidden state mutations that complicate oracle construction.

\subsubsection{Branching Complexity (BC)}

Structural complexity has long been recognized as a factor affecting software testability and maintainability~\cite{mccabe:complexity,mouchawrab2005measurement,bruntink:testability,badri2011empirical}. 

Branching complexity captures the structural difficulty of exercising execution paths during testing. 
It combines the number of control-flow constructs with nesting depth:

\vspace{-14pt}

\begin{small}

\begin{multline*}
BC = (\#if + \#elseIf + \#switch + \#ternary \\
     + \#tryCatch + \#loops) + nestingDepth
\end{multline*}

\end{small}


We compute $BC$ by counting control-flow constructs and tracking the maximum nesting depth during AST traversal.
The metric is inspired by cyclomatic complexity~\cite{mccabe:complexity} but incorporates nesting depth to better reflect reasoning difficulty in practice.

\subsubsection{Asynchronous and Event-Driven Complexity (AC, EC)}
\label{subsubsec:async-event-complexity}

Prior work has adapted testability concepts to concurrent systems through metrics capturing synchronization and shared-state interactions~\cite{yu2016predicting}.
We extend this line of work to modern \js ecosystems by modelling promise-based coordination, timers, event emitters, framework-level handlers, and other event-driven execution patterns.

To capture structural characteristics associated with testing challenges in modern \js systems, we introduce two additional dimensions reflecting asynchronous coordination and event-driven execution.

\vspace{-12pt}

\begin{small}

\begin{multline*}
AC = \#await + \#thenCatchFinally \\
     + \#PromiseAllRaceAny + \#timers + chainDepth
\end{multline*}

\end{small}

\vspace{-20pt}

\begin{small}

\begin{multline*}
EC = \#domListeners + \#eventEmitterUses \\
     + \#frameworkHandlers
\end{multline*}

\end{small}

These metrics quantify coordination complexity introduced by asynchronous constructs and event registration patterns, which complicate test orchestration and observation.

\subsubsection{Encapsulation Level (EL)}
\label{subsubsec:encapsulation}

Encapsulation and hidden state have long been recognized as factors influencing software testability~\cite{binder:oo,jungmayr:testability}. We define Encapsulation Level (EL) as the degree to which a function interacts with hidden or externally scoped state.

\vspace{-14pt}

\begin{small}

\begin{multline*}
EL = \#closureReads + \#closureWrites \\
     + \#moduleSingletonAccess + \#privateStateHeuristics
\end{multline*}

\end{small}

We compute $EL$ by identifying free variables, writes to module-level objects, and accesses to private or conventionally private fields.

Hidden state reduces test isolation and increases setup complexity. 
This metric adapts traditional encapsulation concepts~\cite{binder:oo,jungmayr:testability} to \js flexible scoping rules.

\subsubsection{Side-Effect Index (SEI)}
\label{subsubsec:side-effects}

Side effects can reduce controllability and observability by introducing hidden environmental dependencies, complicating test isolation, and increasing test fragility. 
We quantify their density relative to function size:

\begin{small}

\[
SEI = \frac{\#sideEffectCalls}{LOC_f}
\]

\end{small}

Side-effect calls include known I/O operations, timers, randomness, and environment-dependent APIs.

\subsubsection{Composite Testability Score (CTS)}
\label{subsubsec:cts}

CTS is intended to support comparative characterization and aggregation across functions, files, and projects, rather than serve as a predictive model of testing effectiveness. Accordingly, it should be interpreted as an operational structural indicator rather than an absolute measure of testing difficulty.

Composite indicators have been widely used in software quality assessment, including maintainability indices, quality assessment models, and measurement frameworks that aggregate multiple quality dimensions into a single summary indicator~\cite{freedman:testability,oman1994construction,garousi:survey}. 
Following this tradition, we aggregate the seven normalized dimensions into CTS to provide a concise characterization of structural testability while preserving the underlying dimensions for interpretation.




After normalization to $[0,1]$ and consistent directionality, CTS is defined as the unweighted mean of the dimensions:

\vspace{-12pt}

\begin{small}

\begin{multline*}
CTS = \widehat{C} + \widehat{O}
      + (1-\widehat{BC}) + (1-\widehat{AC})\\
      + (1-\widehat{EC}) + (1-\widehat{EL})
      + (1-\widehat{SEI})
\end{multline*}

\end{small}

We employ equal inter-metric weighting to avoid imposing subjective assumptions regarding the relative importance of different structural testability dimensions.



\subsection{Intra-Metric Weights}
\label{subsec:weights}

While CTS combines all seven dimensions equally, individual metrics use non-uniform intra-metric weights to reflect the relative importance of their constituent structural signals. These weights are derived from prior testability and testing literature, supplemented by pilot analyses on representative JavaScript projects. Full weight configurations, rationales, and sensitivity analyses are available in the artifact.

In general, signals that increase hidden dependencies or coordination complexity receive stronger penalties, while signals that improve controllability or observability receive higher positive weights. For example, explicit parameters contribute more strongly to controllability than default parameters~\cite{freedman:testability,binder:oo}; return values contribute more strongly to observability than indirect effects~\cite{zhang:assertions}; asynchronous combinators and timers receive higher penalties than linear asynchronous chains due to their increased coordination complexity~\cite{alimadadi:promises,turcotte:drasync}; and writes to hidden state are penalized more heavily than reads because they complicate test isolation and oracle construction~\cite{binder:oo,jungmayr:testability}.
%

\subsection{Metric Scope and Limitations}
\label{subsec:metric-discussion}

These metrics rely on static analysis and therefore approximate dynamic behaviour. 
Certain runtime-dependent phenomena (such as dynamic dispatch, runtime code generation, or data-dependent execution paths) may not be fully captured.

In addition, some metrics rely on heuristics (e.g., identifying side-effect APIs or event handlers), which may under- or over-approximate behaviour in specific cases. 
Consequently, metric values should be interpreted as comparative structural indicators of testability characteristics rather than absolute measures of real-world testing effectiveness.

\section{Static Analysis Framework for Structural Testability Analysis}
\label{sec:approach}

This section describes the static analysis framework used to operationalize the structural testability dimensions introduced in \autoref{sec:metrics}. The framework extracts language-aware structural entities corresponding to controllability, observability, complexity, encapsulation, side effects, asynchronous coordination, and event-driven execution. Consistent with the goals of this study, the analysis relies solely on static information and does not require program execution, instrumentation, test suites, or runtime traces, enabling scalable and uniform characterization across diverse \js systems.

\subsection{Analysis Pipeline Overview}
\label{subsec:pipeline}

The analysis proceeds through a deterministic pipeline. Given a project directory, our approach first performs source discovery using configurable glob patterns. Executable \js files (e.g., \texttt{.js}, \texttt{.mjs}, \texttt{.cjs}) are included, while generated artifacts, dependency directories (e.g., \texttt{node\_modules}), build outputs, coverage artifacts, and minified files are excluded. This step produces a project-specific analyzable file set.

Each file is parsed into an abstract syntax tree (AST), configured to support modern ECMAScript features such as JSX, optional chaining, dynamic imports, private fields, and class properties. Parsing failures are logged and excluded.

Next, our technique identifies all function-like constructs, including function declarations, function expressions, arrow functions, and class methods. Each function is assigned a stable identifier composed of the project identifier, file path, function name (if available), and source location. This identifier enables traceability across aggregation levels and ensures consistent measurement across repeated analyses.

\subsection{Function-Level Structural Analysis}
\label{subsec:function-analysis}

Functions constitute the primary unit of analysis. For each function, we perform a localized AST traversal.
Traversals are restricted to the function body and do not require inter-procedural analysis, resulting in linear-time complexity with respect to AST size.

During traversal, the analysis extracts structural features corresponding to the testability dimensions defined in \autoref{sec:metrics}, including controllability, observability, control-flow complexity, asynchronous coordination, event-driven behaviour, encapsulation, and side effects.

Scope inspection distinguishes between locally declared variables, closure variables, imported bindings, and global references. This enables detection of hidden state access and external dependencies without requiring speculative type inference.
Asynchronous constructs such as \code{async}/\code{await}, promise chains, timers, and combinators are detected syntactically. Event-driven patterns, including listener registration and framework-level route handlers, are similarly identified through AST pattern matching. 
All detected structural signals are recorded as raw feature counts at the function level. At this stage, metrics remain independent and no normalization or cross-dimensional aggregation is performed.

\subsection{Metric Construction}
\label{subsec:metric-construction}

Raw structural features are aggregated within each dimension using the weighting schemes described in \autoref{sec:metrics}. These weights operationalize established testability principles: signals that improve controllability or observability contribute positively, whereas signals associated with hidden dependencies, coordination complexity, encapsulation barriers, or environmental interactions contribute negatively. Metrics are constructed independently at this stage, preserving the interpretability of individual dimensions prior to normalization and aggregation.

\subsection{Normalization and Composite Scoring}
\label{subsec:approach-normalization}

Raw metric distributions vary substantially in scale and exhibit heavy-tailed behaviour. To enable cross-metric comparison, we apply dataset-wide robust normalization.
For each metric $M$, we compute the 5th and 95th percentiles across all analyzed functions. Raw values are Winsorized by clamping to this interval to reduce sensitivity to extreme outliers:

\vspace{-4pt}

\begin{small}

\[
M_{clamped} = \min(\max(M, P_5(M)), P_{95}(M))
\]

\end{small}

The clamped values are then scaled to the unit interval:

\vspace{-2pt}

\begin{small}

\[
\hat{M} =
\frac{M_{clamped} - P_5(M)}
     {P_{95}(M) - P_5(M)}
\]

\end{small}

For metrics where larger raw values indicate lower testability (e.g., structural complexity or side effects), normalized values are inverted to ensure consistent directionality. After alignment, higher normalized values uniformly indicate higher testability.

The composite testability score (CTS) for a function $f$ is defined as the unweighted mean of its normalized metric dimensions:

\vspace{-14pt}

\begin{small}

\[
\text{CTS}_f =
\frac{1}{k}
\sum_{i=1}^{k} \hat{M}_i(f)
\]

\end{small}

where $k$ denotes the number of metric dimensions. 
Consistent with the role of CTS as a comparative structural indicator, we adopt equal inter-metric weighting to avoid imposing subjective assumptions about the relative importance of different testability dimensions. The underlying dimensions remain available for interpretation throughout the analysis.

CTS is intended as a comparative structural indicator that summarizes multiple testability dimensions for characterization and aggregation. It is not intended to serve as a predictive model of testing effort or testing effectiveness.

\subsection{Aggregation Across Granularity Levels}
\label{subsec:aggregation-approach}

Although metrics are computed at the function level, the approach supports hierarchical aggregation.
At the file level, we compute robust summary statistics including median CTS, interquartile range (IQR), and the proportion of low-testability functions. At the project level, analogous statistics are computed across all functions within a repository.
This multi-level aggregation enables distributional characterization of testability (RQ1),
analysis of structural characteristics and archetypes of low-CTS regions (RQ2),
and ecosystem-level comparisons across projects (RQ3).

\subsection{Implementation}
\label{subsec:implementation}

The framework is implemented in Node.js/TypeScript and relies on \code{@babel/parser} and \code{@babel/traverse} for AST construction and traversal~\cite{babel-parser-docs,babel-traverse-docs}. 
The implementation exports machine-readable artifacts (JSON/CSV) at function, file, and project levels, enabling subsequent statistical analysis.
An open-source version of our analysis pipeline, experimental artifacts, and data is available.\footnote{\url{https://github.com/SEatSFU/JSTestabilityStudy}}

\section{Empirical Study Design and Methodology}
\label{sec:study-design}

This section describes the design of our empirical study of structural testability in \js projects.
Building on the metrics defined in \autoref{sec:metrics} and the static analysis pipeline in \autoref{sec:approach}, we present the research questions, subject systems, data collection procedure, operational definitions, and analysis strategy.
Our study is observational and cross-sectional: all measurements are derived from static analysis of fixed project revisions.

\begin{table*}[t]
\centering
\renewcommand{\arraystretch}{0.9}
\resizebox{\textwidth}{!}{
\begin{tabular}{l l r r r r r r r r r r r r p{2cm} r}
\toprule
Project & Domain & LOC & Stars & Commits & Issues & \makecell{Age\\(Y)} & \makecell{\#\\Files} & \makecell{\#\\Funs} & \makecell{Med\\CTS} & \makecell{Mean\\CTS} & \makecell{IQR\\CTS} & \makecell{\%\\Low} & \makecell{\#\\Low} & \makecell{Testing\\ Framework} & \# Tests \\
\midrule
\href{https://github.com/expressjs/express}{1. Express} & Web Framework & 39668 & 68791 & 6130 & 187 & 17 & 6 & 109 & 0.91 & 0.86 & 0.14 & 10.09 & 11 & mocha, supertest & 1122 \\
\href{https://github.com/restify/node-restify}{2. node-restify} & Web Framework & 45864 & 10713 & 1727 & 130 & 15 & 53 & 347 & 0.83 & 0.81 & 0.20 & 23.92 & 83 & mocha, chai, sinon & 208 \\
\href{https://github.com/solidjs/solid}{3. solid} & UI Framework & 57500 & 35211 & 1864 & 132 & 8 & 30 & 522 & 0.81 & 0.79 & 0.10 & 20.50 & 107 & vitest & 450 \\
\href{https://github.com/lodash/lodash}{4. Lodash} & Utility Library & 91116 & 61627 & 7688 & 112 & 14 & 16 & 815 & 0.93 & 0.89 & 0.10 & 6.75 & 55 & qunit & 2513 \\
\href{https://github.com/honojs/hono}{5. hono} & Web Framework & 111046 & 28938 & 2539 & 355 & 4 & 272 & 7326 & 0.77 & 0.74 & 0.18 & 41.69 & 3054 & vitest & 2306 \\
\href{https://github.com/mochajs/mocha}{6. mocha} & Testing & 116430 & 22874 & 3919 & 227 & 15 & 61 & 781 & 0.82 & 0.80 & 0.20 & 24.84 & 194 & mocha, karma, chai, sinon & 1608 \\
\href{https://github.com/preactjs/preact}{7. preact} & UI Framework & 116952 & 38402 & 6092 & 154 & 10 & 79 & 536 & 0.82 & 0.80 & 0.14 & 21.08 & 113 & playwright, vitest & 1218 \\
\href{https://github.com/fastify/fastify}{8. fastify} & Web Framework & 155134 & 35661 & 4636 & 108 & 9 & 32 & 472 & 0.82 & 0.80 & 0.19 & 22.25 & 105 & tap & 1607 \\
\href{https://github.com/colinhacks/zod}{9. zod} & Utility Library & 170920 & 41917 & 2809 & 247 & 6 & 89 & 1971 & 0.83 & 0.83 & 0.11 & 12.94 & 255 & vitest & 1733 \\
\href{https://github.com/vitejs/vite}{10. Vite} & Build \& Bundler & 224454 & 78333 & 8966 & 627 & 6 & 461 & 3482 & 0.82 & 0.80 & 0.15 & 22.09 & 769 & vitest & 1504 \\
\href{https://github.com/nestjs/nest}{11. nest} & Web Framework & 258308 & 74705 & 20024 & 49 & 9 & 940 & 6254 & 0.80 & 0.78 & 0.13 & 24.70 & 1545 & jest & 3112 \\
\href{https://github.com/vuejs/core}{12. core} & ORM / Data Layer & 281072 & 53035 & 6930 & 1009 & 8 & 256 & 2833 & 0.80 & 0.78 & 0.11 & 24.21 & 686 & vitest & 3338 \\
\href{https://github.com/jestjs/jest}{13. Jest} & Testing & 311578 & 45297 & 7546 & 239 & 12 & 581 & 3899 & 0.82 & 0.80 & 0.12 & 19.21 & 749 & jest & 4580 \\
\href{https://github.com/parcel-bundler/parcel}{14. parcel} & Build \& Bundler & 348292 & 44041 & 3499 & 586 & 9 & 414 & 4075 & 0.82 & 0.80 & 0.15 & 21.55 & 878 & mocha, sinon & 2204 \\
\href{https://github.com/prettier/prettier}{15. prettier} & Developer Tooling & 355738 & 51678 & 11084 & 1435 & 9 & 513 & 2579 & 0.83 & 0.83 & 0.12 & 12.14 & 313 & jest & 600 \\
\href{https://github.com/vitest-dev/vitest}{16. vitest} & Testing & 362776 & 15987 & 5488 & 358 & 4 & 413 & 5004 & 0.82 & 0.80 & 0.15 & 19.68 & 985 & playwright, vitest & 2814 \\
\href{https://github.com/sveltejs/svelte}{17. Svelte} & UI Framework & 393328 & 85898 & 10982 & 965 & 9 & 381 & 3594 & 0.82 & 0.81 & 0.09 & 18.50 & 665 & playwright, vitest & 4684 \\
\href{https://github.com/evanw/esbuild}{18. esbuild} & Build \& Bundler & 410368 & 39796 & 4393 & 597 & 10 & 35 & 3645 & 0.80 & 0.76 & 0.12 & 28.45 & 1037 & Non-JS (Go) & N/A \\
\href{https://github.com/webpack/webpack}{19. Webpack} & Build \& Bundler & 537082 & 66029 & 18403 & 210 & 14 & 624 & 7977 & 0.83 & 0.82 & 0.12 & 13.74 & 1096 & jest & 166 \\
\href{https://github.com/prisma/prisma}{20. Prisma} & ORM / Data Layer & 588460 & 45369 & 12142 & 2483 & 7 & 829 & 6567 & 0.82 & 0.80 & 0.16 & 22.69 & 1490 & jest, vitest & 2498 \\
\href{https://github.com/rollup/rollup}{21. rollup} & Build \& Bundler & 601900 & 26234 & 6184 & 601 & 11 & 290 & 2373 & 0.83 & 0.83 & 0.13 & 12.43 & 295 & mocha & 169 \\
\href{https://github.com/reduxjs/redux}{22. redux} & Utility Library & 816990 & 61444 & 4145 & 43 & 11 & 13 & 53 & 0.83 & 0.83 & 0.13 & 11.32 & 6 & vitest & 193 \\
\href{https://github.com/microsoft/playwright}{23. Playwright} & Testing & 1011026 & 82896 & 16190 & 613 & 6 & 614 & 12765 & 0.81 & 0.78 & 0.13 & 25.19 & 3216 & playwright & 5878 \\
\href{https://github.com/eslint/eslint}{24. ESLint} & Developer Tooling & 1024616 & 27099 & 10620 & 104 & 13 & 412 & 4296 & 0.83 & 0.84 & 0.14 & 9.87 & 424 & mocha + RuleTester & 663 \\
\href{https://github.com/cypress-io/cypress}{25. cypress} & Testing & 1307050 & 49574 & 22683 & 1241 & 11 & 2629 & 46913 & 0.79 & 0.77 & 0.14 & 29.19 & 13693 & cypress, jest, mocha, vitest, chai, sinon & 7524 \\
\href{https://github.com/babel/babel}{26. Babel} & Build \& Bundler & 1329646 & 43886 & 18381 & 761 & 11 & 758 & 6995 & 0.82 & 0.82 & 0.13 & 16.61 & 1162 & jest & 2524 \\
\href{https://github.com/angular/angular}{27. Angular} & UI Framework & 2312504 & 100000 & 36505 & 1099 & 11 & 2150 & 20463 & 0.82 & 0.81 & 0.13 & 19.69 & 4030 & cypress, jasmine, jest, karma & 3303 \\
\href{https://github.com/vercel/next.js}{28. next.js} & UI Framework & 4068178 & 137891 & 32872 & 3372 & 9 & 2291 & 91322 & 0.83 & 0.82 & 0.17 & 19.04 & 17392 & cypress, jest, playwright, vitest & 9932 \\
\href{https://github.com/microsoft/TypeScript}{29. TypeScript} & Developer Tooling & 7193454 & 107898 & 36730 & 5467 & 12 & 539 & 19698 & 0.80 & 0.79 & 0.13 & 25.23 & 4969 & mocha & 1682 \\
\href{https://github.com/swc-project/swc}{30. swc} & Build \& Bundler & 17851380 & 33237 & 11573 & 415 & 8 & 279 & 69328 & 0.88 & 0.86 & 0.15 & 11.69 & 8102 & jest, mocha & 732 \\
\midrule
Total & - & 42492830 & 1614461 & 342744 & 23926 & - & 16060 & 336994 & - & - & - & - & 67479 & - & 70865 \\
Mean  & - & 1416427.67 & 53815.37 & 11424.8 & 797.53 & - & 535.33 & 11233.13 & 0.83 & 0.81 & 0.14 & 19.71 & - & - & 2443.62 \\
\bottomrule
\end{tabular}
}
\caption{Project characteristics and project-level testability summaries. Med/Mean CTS denote the median/mean composite testability score; IQR CTS denotes the interquartile range; \% Low-Test denotes the percentage of functions classified as low-testability using the global threshold.}
\label{tab:benchmarks}
\end{table*}

\subsection{Research Questions}
\label{subsec:rqs}

Our study is guided by the following research questions:

\begin{itemize}
  \item \textbf{RQ1:} What is the distribution of testability across \js projects in practice?
  \item \textbf{RQ2:} What structural profiles emerge among functions with low structural testability?
  \item \textbf{RQ3:} How do project characteristics and testing practices relate to testability in \js projects?
\end{itemize}

RQ1 characterizes the empirical distribution of testability at function, file, and project levels.  
RQ2 investigates the structural characteristics of low-CTS functions and identifies recurring structural archetypes.
RQ3 examines whether project-level characteristics are associated with structural testability.

\subsection{Subject Projects}
\label{subsec:subjects}

We analyze \numOfProjects open-source \js projects selected from GitHub. To ensure non-trivial and actively-maintained subject systems, we require repositories to contain executable \js source files (\texttt{.js}, \texttt{.mjs}, or \texttt{.cjs}), exhibit a coherent project structure, and satisfy minimum thresholds for size ($>$39K LOC) and activity ($>$1,500 commits).
The resulting dataset spans a diverse set of project domains, including developer tooling, UI frameworks, build systems, testing frameworks, and utility libraries. This diversity arises naturally from the composition of popular and actively maintained \js repositories rather than from explicit selection for particular architectural styles or application categories.
In total, it contains \numOfFiles source files, \numOfFunctions functions, and \totalLOC lines of code. \autoref{tab:benchmarks} summarizes the included projects and their metadata.
For each project, we also collect ecosystem metadata including LOC, number of analyzed files and functions, age, GitHub stars, commits, issues, and testing frameworks. Generated artifacts, dependencies, coverage outputs, and non-source assets are excluded uniformly across projects.

\subsection{Data Collection Procedure}
\label{subsec:data-collection}

For each project, we apply the static analysis pipeline described in \autoref{sec:approach}, which performs source discovery and filtering, AST parsing, function identification, and function-level metric computation. Raw feature counts and derived metric values are recorded for every identified function. Parsing failures and coverage statistics are logged to quantify completeness. All projects are analyzed using the same tool version and configuration.
%
%
Metric normalization and CTS construction follow the procedure described in \autoref{subsec:approach-normalization}.

\subsection{Units of Analysis and Aggregation}
\label{subsec:aggregation}

The primary unit of measurement is the function. For RQ1 and RQ2, we analyze function-level structural testability directly. For RQ1 and RQ3, we also aggregate function-level measurements to higher levels:

\begin{itemize}
    \item File level: median CTS, interquartile range (IQR), and proportion of low-CTS functions within each file.
    \item Project level: summary statistics and dimension means aggregated across all functions in a project.
\end{itemize}

This multi-level design allows us to study both localized structural difficulty and broader ecosystem-level patterns.

\subsection{Operational Definition of Low Testability}
\label{subsec:low-testability}


To identify structurally-difficult code regions and support their comparative analysis across projects, we define low-CTS functions using a global percentile-based threshold over the corpus-wide CTS distribution. Let $\text{CTS}_f$ denote the composite testability score of function $f$ in the corpus $\mathcal{D}$. We compute a single threshold $\tau$ as the 20th percentile of all function-level CTS values in $\mathcal{D}$, and classify a function as low-CTS if $\text{CTS}_f \le \tau$.
This definition labels approximately the bottom 20\% of functions in the corpus as low-CTS. Using a global threshold enables consistent cross-project comparison and avoids project-specific cutoffs.

\subsection{Analysis Methodology}
\label{subsec:analysis-methods}

Because several metric and project-level variables exhibit non-normal and heavy-tailed distributions, we employ non-parametric and robust statistical procedures throughout the study. We report effect sizes alongside significance tests and interpret findings conservatively.

\paragraph{RQ1: Distribution of Structural Testability}
We characterize the distribution of CTS at function, file, and project levels using robust descriptive statistics and visualizations.

\paragraph{RQ2: Structural Characteristics of Low-CTS Functions}
We analyze low-CTS functions using two complementary views. First, we compare the structural characteristics of low-CTS and non-low-CTS functions across the testability dimensions. Second, we cluster low-CTS functions to identify recurring structural archetypes and investigate how testability limitations manifest in practice.
To compare structural characteristics between low-CTS and non-low-CTS functions, we use the two-sided Mann--Whitney U test and report Cliff's $\delta$ effect sizes. To identify recurring structural profiles, we apply $k$-means clustering to normalized metric vectors, selecting the number of clusters using silhouette analysis.

\paragraph{RQ3: Ecosystem-Level Associations}
We analyze project-level relationships between structural testability and project characteristics such as size, age, activity, popularity, and testing practices. This analysis includes bivariate associations, framework-related comparisons, and multivariate project models.
For project-level modelling, we use linear models with robust HC3 standard errors. Predictors include project size, activity, and testing intensity, and standardized coefficients are reported to facilitate comparison across predictors.

\subsection{Reproducibility and Quality Controls}
\label{subsec:reproducibility}

To support reproducibility, we fix analyzed commit hashes, record tool versions and configurations, log parsing coverage and exclusions, version output schemas, and export machine-readable artifacts (JSON/CSV). These steps enable independent replication and consistent re-analysis of the study.

\section{RQ1: Distribution of Structural Testability in \js}
\label{sec:rq1-results}

RQ1 asks: \emph{What is the distribution of structural testability across \js projects in practice?}
We examine the Composite Testability Score (CTS) at three levels of granularity: projects, files, and functions.

\subsection{Analysis Method}

To characterize structural testability, we analyze the distribution of CTS across the corpus and aggregate results at the project, file, and function levels. 
At the project level, we summarize CTS using median and interquartile range (IQR) to capture both central tendency and dispersion.  
At the file level, we analyze the concentration of low-CTS functions within individual files.  
At the function level, we examine the global CTS distribution to identify overall structural patterns.

\subsection{Results and Discussion}

\paragraph{Project-Level Testability Profiles}
\begin{figure}[t]
    \centering
    \includegraphics[width=0.450\textwidth]{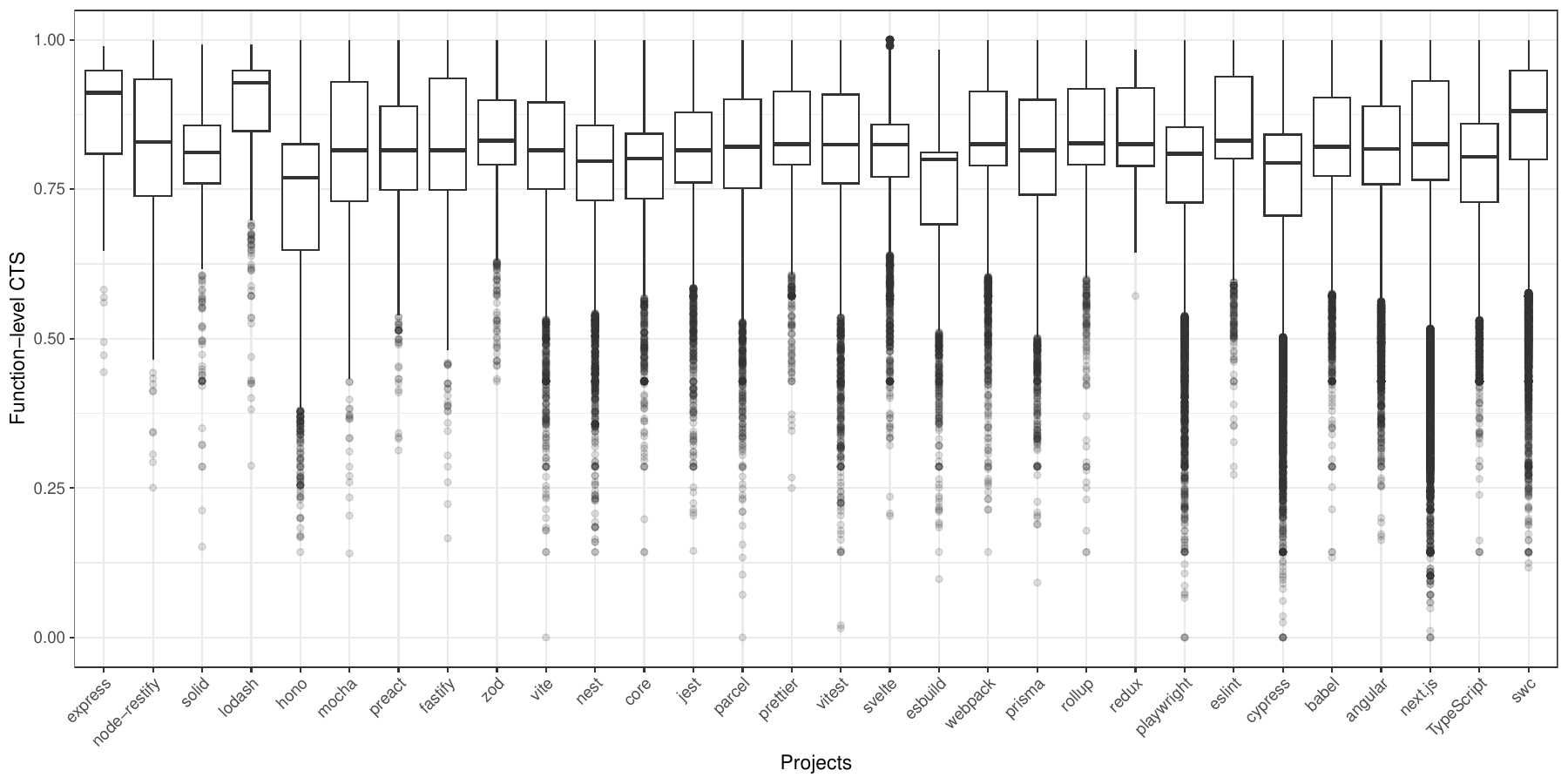}
    \caption{Function-level CTS distributions by project.} 
    \label{fig:project-cts-boxplot}
\end{figure}

Median CTS values across projects range from \minMedianCTS (\textit{Hono}) to \maxMedianCTS (\textit{Lodash}), with most projects clustered between 0.74 and 0.80 (\autoref{tab:benchmarks}, column 10). 
Despite substantial variation in project size, age, and domain, project-level CTS distributions exhibit remarkably similar central tendencies.
However, within-project dispersion is substantial. Interquartile ranges vary between 0.09 and 0.19 (median \medianIQR), indicating that even high-CTS projects contain functions with significantly lower structural testability (column 12). This heterogeneity is visible in \autoref{fig:project-cts-boxplot}, where large projects such as \textit{TypeScript} and \textit{Prettier} exhibit wide CTS distributions despite high medians.

\begin{figure}[t]
    \centering
    \includegraphics[width=0.450\textwidth]{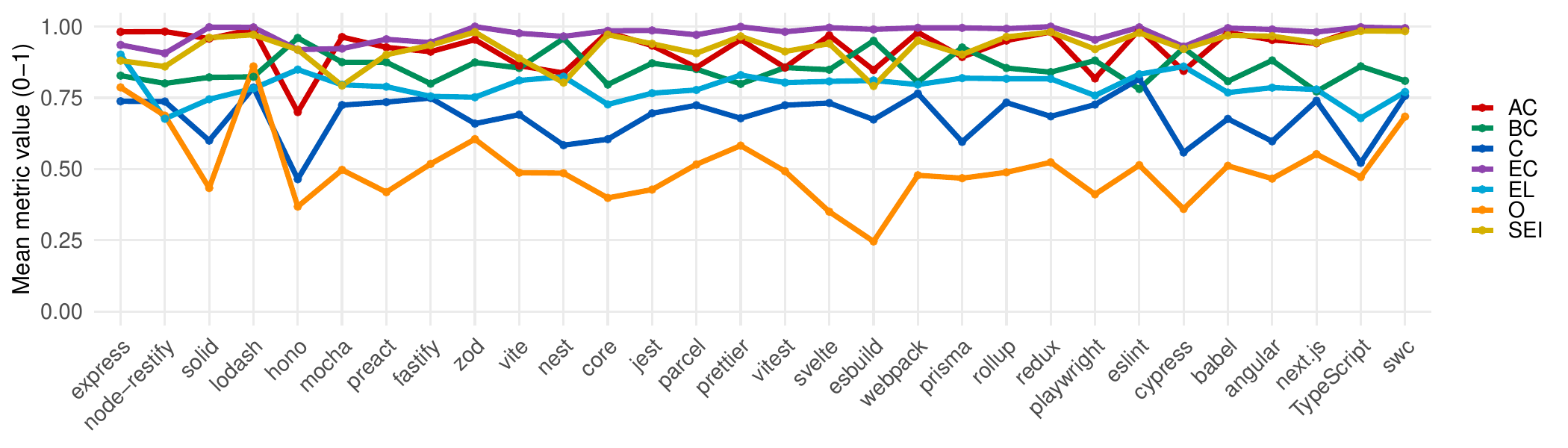}
    \caption{Mean normalized testability dimensions by project, ordered by increasing mean CTS.}
    \label{fig:project-metrics-lines}
\end{figure}

Dimension-level profiles further reveal that similar aggregate CTS values can arise from structurally different combinations of testability characteristics (\autoref{fig:project-metrics-lines}).
This observation suggests that structurally-challenging regions may emerge through multiple structural configurations rather than a single dominant pattern

\paragraph{File-Level Concentration}
\begin{figure}[t]
\centering
\includegraphics[width=0.45\textwidth]{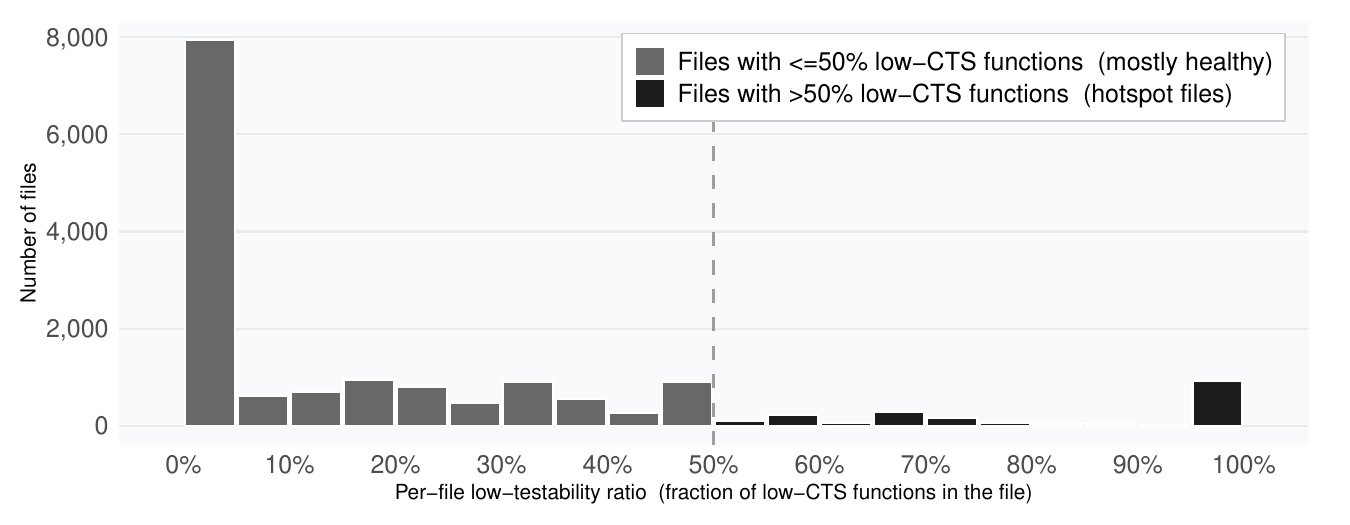}
\caption{
Distribution of low-CTS concentration across files.\\
X-axis: percentage of low-CTS functions in a file;\\
Y-axis: number of files.
}
\label{fig:file-lowtest-dist}
\end{figure}

Low-CTS functions are not evenly distributed across files. \autoref{fig:file-lowtest-dist} shows a strongly right-skewed distribution: most files contain few or no low-CTS functions, while a minority contain dense clusters.

\begin{figure}[t]
\centering
\includegraphics[width=0.45\textwidth]{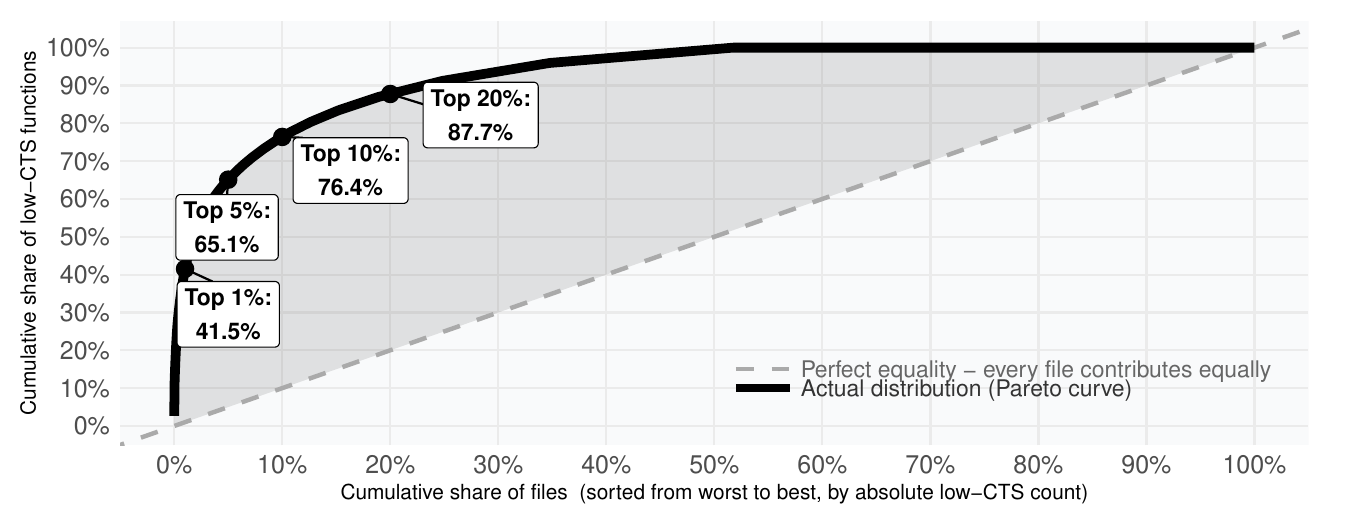}
\caption{
Cumulative concentration of low-CTS functions.\\
X-axis: cumulative percentage of files (sorted by severity);\\
Y-axis: cumulative percentage of low-CTS functions.
}
\label{fig:file-concentration}
\end{figure}

This concentration effect is confirmed by the cumulative distribution in \autoref{fig:file-concentration}, where a small fraction of files accounts for a disproportionately large share of low-CTS functions.
These results indicate that structurally-challenging code tends to be concentrated within specific modules rather than uniformly distributed throughout a system.

\paragraph{Function-Level Distribution}

\begin{figure}[t]
\centering
\includegraphics[width=0.45\textwidth]{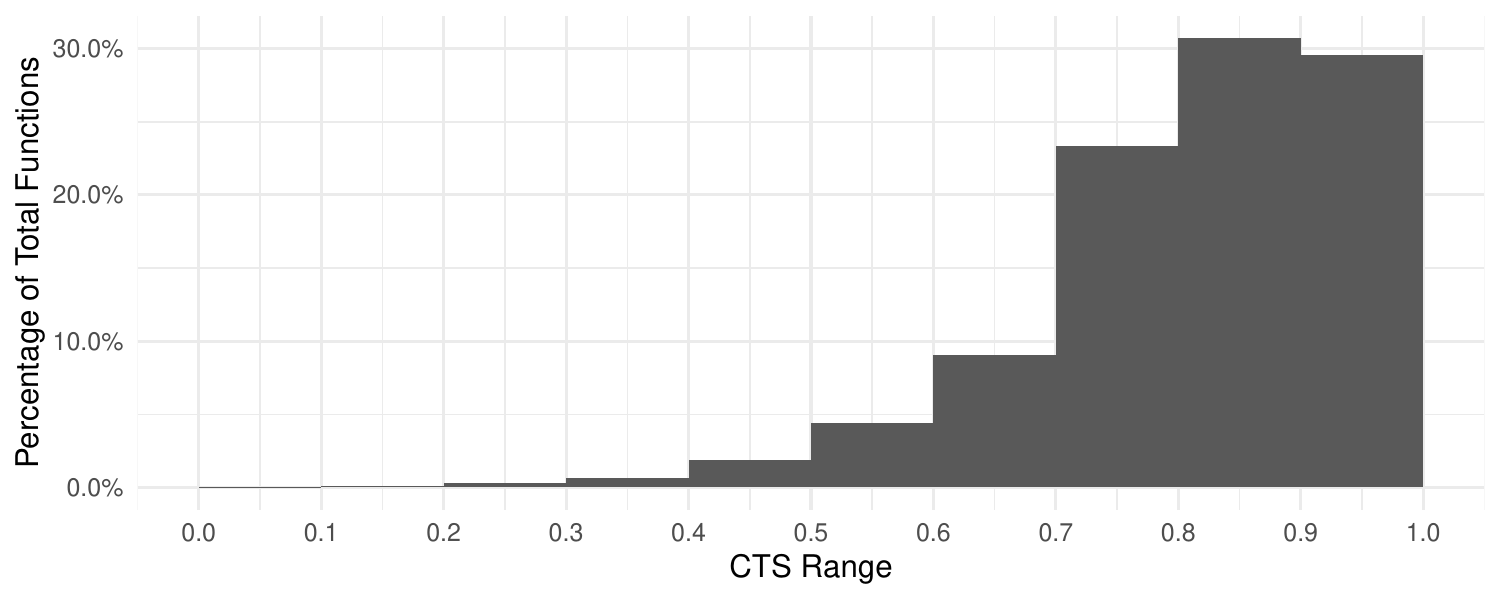}
\caption{Global distribution of composite testability scores (CTS) across all analyzed
functions.}
\label{fig:cts-distribution}
\end{figure}

The global CTS distribution (\autoref{fig:cts-distribution}) exhibits a pronounced lower tail. While most functions occupy the moderate-to-high CTS range (approximately 0.70–0.85), a smaller subset forms a concentrated region of structurally-challenging code. This pattern suggests that structural testability barriers are relatively uncommon but potentially important, motivating the focused analysis of low-CTS functions in RQ2.

Importantly, these low-CTS regions appear even in projects with high median CTS values, reinforcing the observation that structural testability varies substantially within large systems.

\subsection{Summary of Findings}

RQ1 reveals two consistent properties of structural testability in JavaScript systems:

\begin{itemize}
\item Structural heterogeneity: all projects contain both high-CTS and low-CTS functions despite relatively similar project-level CTS distributions..
\item Localized difficulty: low-CTS functions are concentrated within a relatively small subset of files rather than being uniformly distributed across a codebase.
\end{itemize}

Together, these findings suggest that structural testability challenges arise through localized code structures rather than project-wide limitations. We therefore investigate in RQ2 how these low-CTS regions are characterized and whether recurring structural archetypes emerge across projects.





\section{RQ2: Structural Characteristics of Low-CTS Functions}
\label{sec:rq2-results}

RQ2 investigates the structural characteristics of functions with low composite testability scores (CTS). Building on RQ1, we examine how low-CTS functions differ from the remainder of the corpus and whether recurring structural archetypes emerge within these regions.
Columns 13--14 of \autoref{tab:benchmarks} report 
the number and proportion of functions classified as low-CTS.

\subsection{Analysis Method}

We perform two complementary analyses.
First, we compare low-CTS and non-low-CTS functions across the testability dimensions to characterize the structural differences between these regions.
Second, we apply clustering to low-CTS functions to identify recurring structural profiles and investigate how testability limitations manifest in practice.

\subsection{Structural Characteristics of Low-CTS Functions}

\begin{table}[t]
\centering
\caption{Univariate analysis: low vs. non-low functions.}
\label{tab:rq2-univariate}
\renewcommand{\arraystretch}{0.25}
\scriptsize
\begin{tabular}{lcccc}
\toprule
Metric & Median (Low) & Median (High) & Cliff's $\delta$ & $p$-value \\
\midrule
C & 0.47 & 0.77 & -0.59 & $< 1\times10^{-16}$ \\
O & 0.00 & 1.00 & -0.23 & $< 1\times10^{-16}$ \\
BC & 0.79 & 1.00 & -0.39 & $< 1\times10^{-16}$ \\
AC & 1.00 & 1.00 & -0.31 & $< 1\times10^{-16}$ \\
EC & 1.00 & 1.00 & -0.12 & $< 1\times10^{-16}$ \\
EL & 0.53 & 0.94 & -0.62 & $< 1\times10^{-16}$ \\
SEI & 1.00 & 1.00 & -0.23 & $< 1\times10^{-16}$ \\
\bottomrule
\end{tabular}
\end{table}

\autoref{tab:rq2-univariate} summarizes the structural characteristics of low-CTS and non-low-CTS functions across the seven testability dimensions. Low-CTS functions exhibit lower controllability, observability, and encapsulation levels, together with higher complexity-related characteristics.

The largest distribution differences are observed for encapsulation level (EL, $\delta=-0.62$) and controllability (C, $\delta=-0.59$), 
indicating that low-CTS functions more frequently involve hidden-state interactions and limited externally controllable inputs.
Branching complexity (BC) and asynchronous complexity (AC) exhibit moderate differences ($\delta=-0.39$ and $-0.31$), while observability (O) and side-effect index (SEI) show smaller but still meaningful separation. Event-driven complexity (EC) differs only marginally between the two groups ($\delta=-0.12$).

\subsection{Structural Archetypes of Low-Testability Functions}
\label{subsec:rq2-archetypes}

\begin{figure}[t]
\centering
\includegraphics[width=0.45\textwidth]{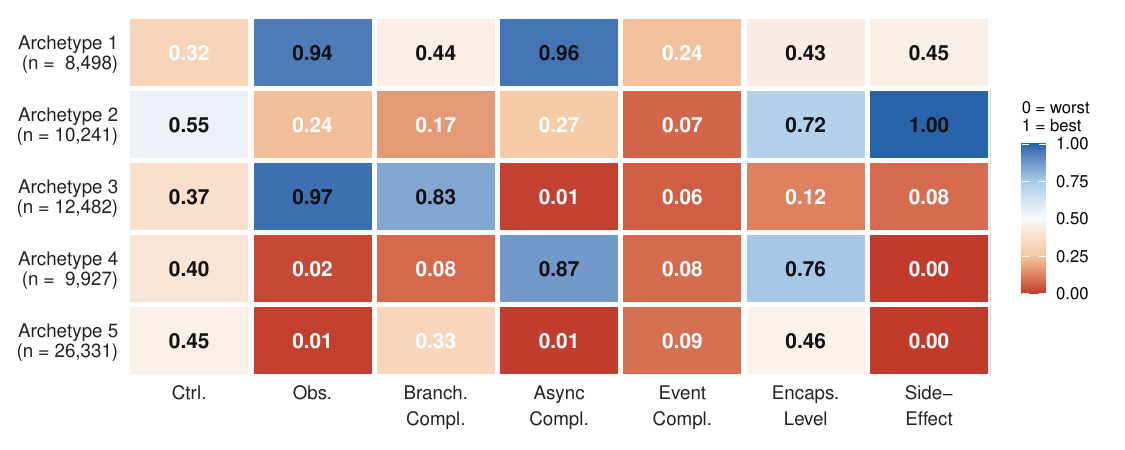}
\caption{Structural archetypes of low-CTS functions.}
\label{fig:rq2-archetypes}
\end{figure}

While the preceding analysis summarizes aggregate structural characteristics, it does not reveal whether low-CTS functions share a common structure or arise through multiple distinct structural configurations.
Applying $k$-means clustering to the 67,479 low-CTS functions yields five archetypes (\autoref{fig:rq2-archetypes}). These clusters reveal distinct structural patterns:

\begin{itemize}
\item (A1) Asynchronous coordination functions: functions with high asynchronous complexity but limited controllability.
\item (A2) Side-effect-dominated functions: functions with extensive side effects and weak observability.
\item (A3) Branching-intensive synchronous functions: highly-observable functions with substantial branching complexity.
\item (A4) Opaque asynchronous functions: async routines with minimal observable outputs.
\item (A5) Structurally-unobservable functions: the largest group, characterized by near-zero observability and minimal external effects.
\end{itemize}

These archetypes demonstrate that low-CTS functions are not structurally homogeneous. Rather than reflecting a single form of testing difficulty, low-CTS regions emerge through multiple recurring structural configurations involving asynchronous coordination, observability limitations, hidden-state interactions, side effects, and branching complexity.
This diversity suggests that different classes of testability barriers may require different testing and refactoring strategies.

\subsection{Summary of Findings}

RQ2 reveals that low-CTS functions differ systematically from the remainder of the corpus across several structural dimensions, particularly controllability, encapsulation, and complexity-related characteristics.
More importantly, low-CTS functions are not structurally homogeneous. Clustering analysis identifies multiple recurring structural archetypes, including asynchronous coordination routines, side-effect-dominated functions, and structurally-unobservable code.

These findings suggest that structurally-challenging regions of \js systems arise through several recurring configurations rather than a single dominant pattern, highlighting the multi-dimensional nature of structural testability.



\begin{figure}[t]
    \centering
    \includegraphics[width=\linewidth]{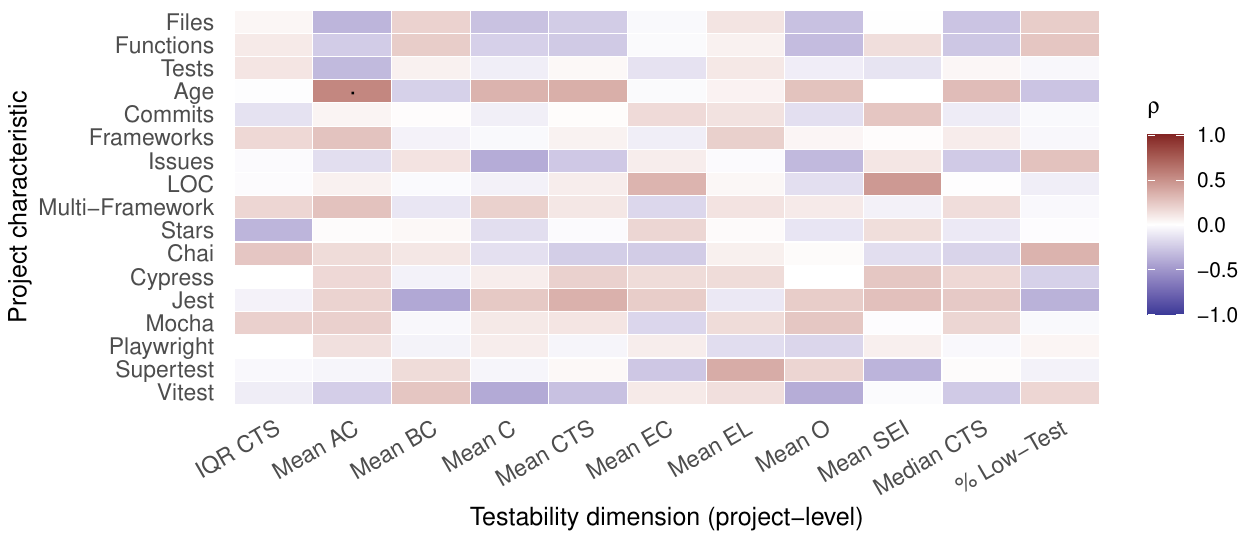}
    \caption{Spearman correlations between project characteristics and testability dimensions. Positive values (red) indicate positive associations.}
    \label{fig:rq3_dims_heatmap}
\end{figure}


\section{RQ3: Project Characteristics and Testability}
\label{sec:rq3-results}

RQ3 asks: \emph{How do project-level ecosystem characteristics relate to structural testability in \js systems?}
While RQ1 and RQ2 focused on function-level distributions and structural characteristics, RQ3 examines whether broader project characteristics are associated with project-level testability patterns.

\subsection{Analysis Method}

We analyze project-level testability using two complementary views.
First, we examine bivariate associations between ecosystem characteristics and aggregated testability measures.
Second, we fit exploratory multivariate models to investigate whether combinations of project size, activity, and testing intensity are associated with project-level testability outcomes.

\subsection{Bivariate Project-Level Associations}


At the bivariate level, project characteristics exhibit generally weak associations with project-level structural testability. None of the primary testability outcomes remain statistically significant after multiple-comparison correction, suggesting that structural testability is not strongly explained by coarse repository characteristics alone.

At the dimension level, several exploratory trends emerge. Older projects tend to exhibit higher asynchronous complexity, while larger projects tend to exhibit greater side-effect intensity. However, most relationships remain weak, and only a small subset survives FDR correction (\autoref{fig:rq3_dims_heatmap}). We also observe no statistically robust association between testing framework choice and project-level testability outcomes.

\subsection{Multivariate Ecosystem Modeling}

To assess combined ecosystem-level associations, we fit project-level linear models using size ($\log(\text{LOC})$), activity (commits per year), and testing intensity (tests per function) as predictors.

\paragraph{Mean structural testability}
\begin{figure}[t]
    \centering
    \includegraphics[width=0.85\linewidth]{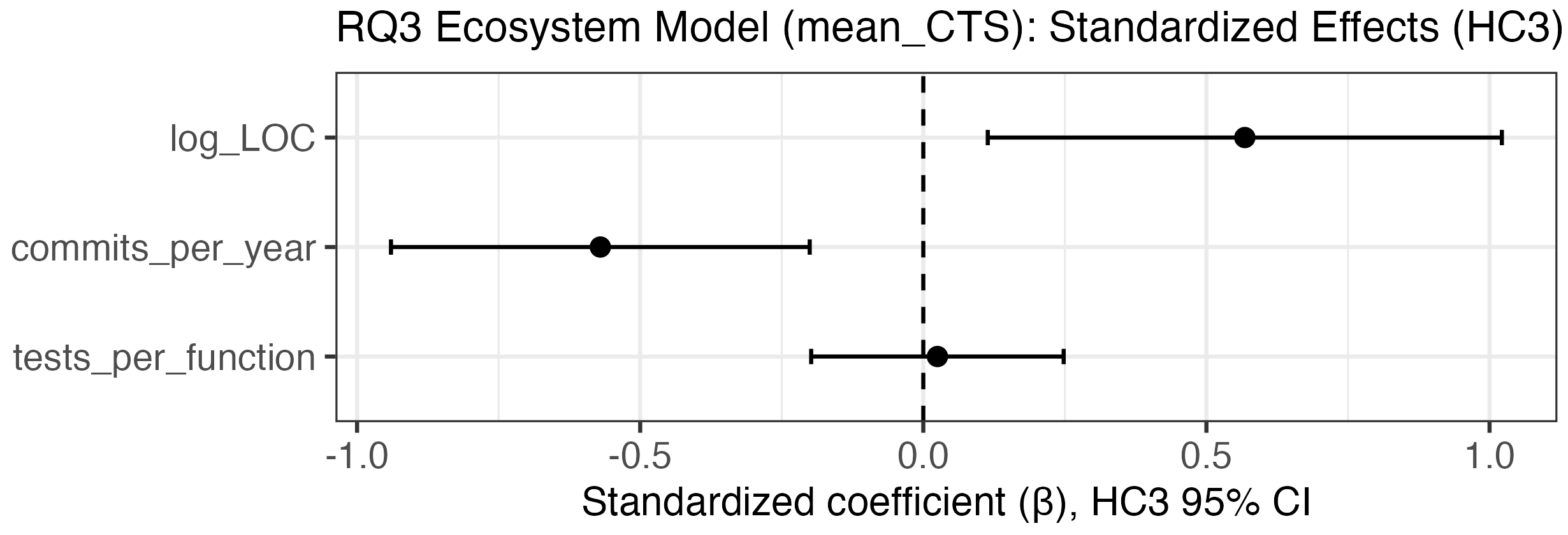}
    \caption{Standardized coefficients ($\beta$, HC3 95\% CI) for the ecosystem model predicting $\textit{mean\_CTS}$.}
    \label{fig:rq3_mean_cts_coef}
\end{figure}



\autoref{fig:rq3_mean_cts_coef} shows the standardized coefficients for the full model predicting mean CTS. Project size is positively associated with mean CTS ($\beta \approx +0.57$, $p<0.05$), whereas ecosystem activity is negatively associated ($\beta \approx -0.57$). Model comparisons indicate that activity provides the largest incremental explanatory gain beyond project size, while testing intensity contributes little additional explanatory power.

\paragraph{Distributional outcomes}

While median CTS remains relatively stable across projects, prevalence-based outcomes exhibit clearer structure. Larger projects tend to contain a smaller proportion of low-CTS functions, whereas more active projects contain a greater proportion.
This pattern suggests that project size and ongoing development activity are associated with structural testability in different ways.

\paragraph{Dimension-level effects}

\begin{figure}[t]
    \centering
    \includegraphics[width=0.9\linewidth]{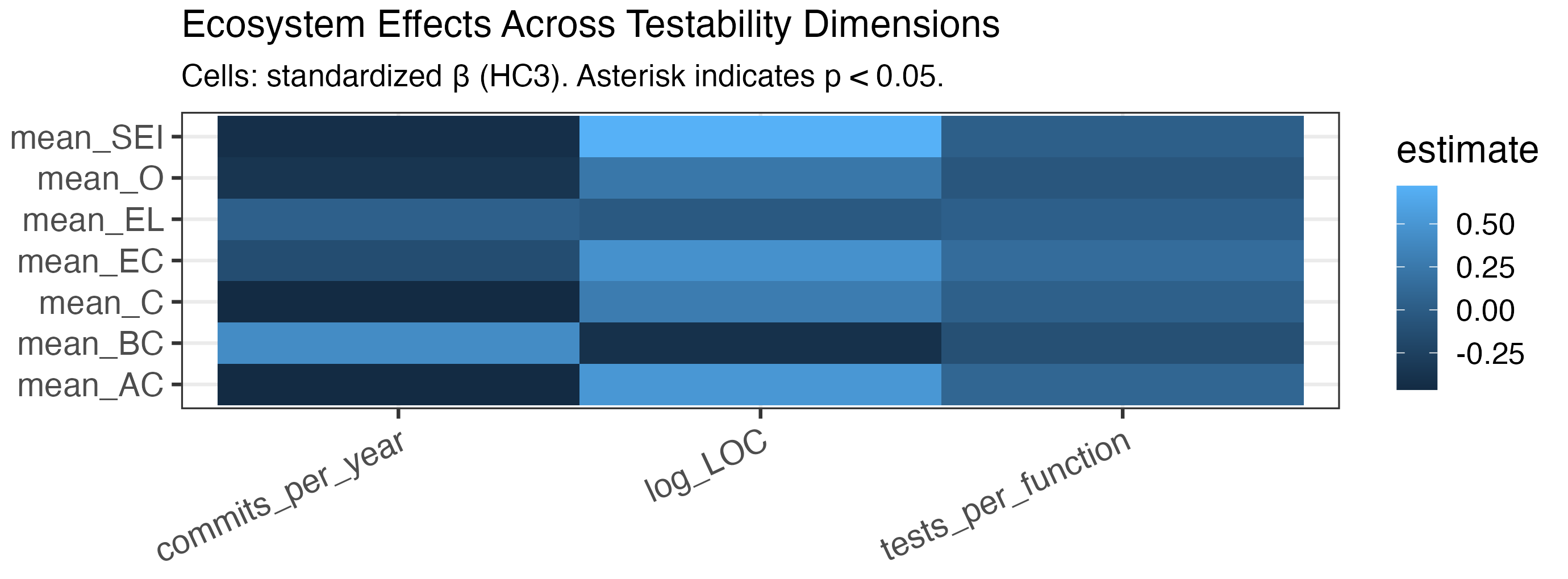}
    \caption{Standardized ecosystem associations ($\beta$, HC3) across structural testability dimensions. Asterisks indicate $p < 0.05$.}
    \label{fig:rq3_dimension_heatmap}
\end{figure}

Cross-dimension models reveal similar directional trends (\autoref{fig:rq3_dimension_heatmap}), although individual estimates remain underpowered. Project size tends to exhibit positive associations across several dimensions, whereas ecosystem activity tends to exhibit negative associations. Testing intensity remains weak throughout.


Taken together, these results suggest that broader project context may be related to the prevalence and distribution of structurally-challenging code. In particular, project size and ecosystem activity exhibit stronger associations with structural testability than age, popularity, or framework choice in this dataset.

\subsection{Summary of Findings}

RQ3 indicates that coarse repository characteristics alone provide limited explanatory power for structural testability patterns.
Among the ecosystem variables considered, project size exhibits the strongest positive association with mean CTS, whereas development activity exhibits the strongest negative association.


These findings complement RQ1 and RQ2 by indicating that structurally challenging regions arise within broader project contexts. Although most ecosystem-level relationships are modest, project activity and growth appear more informative than repository metadata alone when characterizing structural testability patterns.

\section{Threats to Validity}
\label{sec:threats}

Our study measures structural testability using a composite score
derived from seven static metrics. While these metrics capture
structural properties associated with testability, they approximate
rather than directly measure developers’ ability to design effective
tests. Factors such as developer expertise, runtime behaviour, and test
quality are therefore not captured.

All measurements rely on static analysis of \js code, which may miss
dynamic constructs such as runtime-generated functions or reflective
behaviour. To mitigate this risk, we apply a uniform analysis pipeline
across all projects and exclude generated artifacts and dependencies.
Our dataset includes \numOfProjects large open-source \js projects from
GitHub. Although these projects span multiple domains and vary widely in
size and activity, results may not generalize to smaller repositories,
proprietary systems, or other languages.
Finally, some analyses involve multiple statistical comparisons and
project-level models with a moderate sample size. We mitigate this risk
using non-parametric tests, effect sizes, and false discovery rate
(FDR) correction; nevertheless, project-level associations should be
interpreted as exploratory rather than causal.

For reproducibility, we provide the analysis tool, pipeline, scripts,
and dataset in an \href{https://github.com/SEatSFU/JSTestabilityStudy}{artifact repository}.
\section{Related Work}
\label{sec:related-work}


\subsection{Software Testability: Concepts, Metrics, and Studies}

Testability as a software quality attribute traces back to
Freedman~\cite{freedman:testability}, whose controllability and
observability framework motivates our metric dimensions, and to
Voas and Miller~\cite{voas:testability}, who framed testability as a
measurable property affecting fault detection.
Binder~\cite{binder:oo} extended these ideas to object-oriented (OO) design,
linking encapsulation and coupling to unit-level testability.
Empirical studies further show that structural metrics relate to
testability.
Bruntink and van Deursen~\cite{bruntink:testability}
demonstrated that static metrics can meaningfully distinguish
easy-to-test from hard-to-test Java classes, while
Jungmayr~\cite{jungmayr:testability} highlighted how software
dependencies degrade testability. Garousi et
al.~\cite{garousi:survey} consolidate this evidence across 208 papers, and
Terragni et al.~\cite{terragni:testability} show that
quality-normalized test effort correlates more strongly with structural
metrics than raw effort. 
%
However, this body of work focuses almost exclusively on statically-typed OO
languages. To our knowledge, our work is the first large-scale empirical characterization of structural testability in
JavaScript.

\subsection{JavaScript-Specific Testability and Empirical Studies}

A closely-directly related work is by Fard and
Mesbah~\cite{fard:uncovered}, who identified closures, callbacks, and
dynamic DOM manipulation as structural coverage barriers in JavaScript,
motivating our encapsulation~(EL),
event-driven~(EC), and asynchronous~(AC) metrics. Andreasen et
al.~\cite{andreasen:survey} survey broader challenges in \js
analysis and test generation.
Several studies examine asynchronous behaviour in \js.
Alimadadi et al.~\cite{alimadadi:promises}
showed how broken promises produce bugs that evade conventional tests,
and Turcotte et al.~\cite{turcotte:drasync} catalogued asynchronous
anti-patterns that map closely to the failure modes surfaced by our
clustering. Ganji et al.~\cite{ganji:jscope} formalized asynchronous
coverage criteria, motivating our AC and EC dimensions, and
Hashemi et al.~\cite{hashemi:flaky} linked concurrency issues
to test flakiness in large JavaScript projects, grounding our
AC metrics.
~\cite{carstensen:thesis} is the closest
predecessor to our work, adapting OO proxy metrics such as cyclomatic
complexity and LOC to JavaScript. We extend this line of work by introducing metrics
tailored to JavaScript's asynchronous, event-driven, and closure-heavy
nature, operating at the function level, and performing
multivariate modelling and  clustering to characterize
archetypes of low testability.

\subsection{Automated Test Generation for JavaScript}

Arteca et al.'s Nessie~\cite{arteca:nessie} was among the first tools
to target asynchronous callbacks in JavaScript, 
highlighting the difficulty of achieving coverage in asynchronous code. 
Olsthoorn et al.~\cite{olsthoorn:syntest}
and Stallenberg et al.~\cite{stallenberg:guess} proposed search-based
and probabilistic unit test generation approaches with mixed
success on asynchronous behaviour.
Among LLM-based approaches, Sch\"{a}fer et
al.~\cite{schafer:testpilot} (TestPilot) reported strong performance on
API-level testing but weaker results for asynchronous and internal
behaviours.
Finally, Inozemtseva and Holmes~\cite{inozemtseva:coverage} showed that
coverage alone is not strongly correlated with test suite effectiveness,
and Zhang and Mesbah~\cite{zhang:assertions} found assertion
density to be a stronger predictor. These results motivate the need for
structural testability metrics as a complement to coverage.
\section{Concluding Remarks}
\label{sec:conclusion}


This paper presented a large-scale empirical characterization of structural testability in modern \js systems. Our analysis of 30 open-source \js projects comprising \totalLOC LOC reveals that structural testability is highly heterogeneous: low-CTS functions are concentrated within a relatively small subset of files, exhibit distinct structural characteristics, and form multiple recurring archetypes rather than a single dominant pattern.

These findings provide new empirical insight into structural testability in modern \js ecosystems and establish a foundation for future studies of automated testing, testability-aware refactoring, and software quality assessment.

\section*{Acknowledgements}
This research was supported in part by the Natural Sciences and Engineering Research Council of Canada (NSERC). We thank Arshia Lak for his assistance with data collection and result preparation.

\balance

\bibliographystyle{IEEEtran}
\bibliography{IEEEabrv,refs}

\end{document}